\documentclass[aps,prx,twocolumn,10pt,nopacs,floats,floatfix,superscriptaddress]{revtex4-2}
\usepackage{graphicx}
\usepackage{dcolumn}
\usepackage{bm}
\usepackage{amsmath}
\usepackage{amsfonts}
\usepackage{color}
\usepackage{braket}
\usepackage{verbatim}

\usepackage{hyperref}

\definecolor{cardinal}{rgb}{0.6,0,0}
\definecolor{darkgreen}{rgb}{0,0.4,0}
\definecolor{golden}{rgb}{0.92, 0.7, 0}
\definecolor{midnight}{rgb}{0, 0, 0.5}
\definecolor{darkblue}{rgb}{0, 0, 0.7}

\def\he4{$^4$He}
\def\hel3{$^3$He}
\def\Am3{\AA$^{-3}$}
\def\beq{\begin{equation}}
\def\eeq{\end{equation}}

\newcommand{\Rv}{{\bf R}}

\newcommand{\bkfa}{Ba$_{1-x}$K$_x$Fe$_2$As$_2$}

\newcommand{\zh}{{\hat{\bf z}}}

\newcommand{\be}{\begin{equation}}
\newcommand{\ee}{\end{equation}}
\newcommand{\bea}{\begin{eqnarray}}
\newcommand{\eea}{\end{eqnarray}}
\newcommand{\bse}{\begin{subequations}}
\newcommand{\ese}{\end{subequations}}
\newcommand{\upd}{\text{d}}

\begin{document}

\title{Borromean supercounterfluids at finite temperatures}

\author{Alexandru Golic} 
\affiliation{Department of Physics, The Royal Institute of Technology, Stockholm SE-10691, Sweden}

\author{Igor Timoshuk} 
\affiliation{Department of Physics, The Royal Institute of Technology, Stockholm SE-10691, Sweden}
\affiliation{Wallenberg Initiative Materials Science for Sustainability, Department of Physics, The Royal Institute of Technology, Stockholm SE-10691, Sweden}

\author{Egor Babaev} 
\affiliation{Department of Physics, The Royal Institute of Technology, Stockholm SE-10691, Sweden}
\affiliation{Wallenberg Initiative Materials Science for Sustainability, Department of Physics, The Royal Institute of Technology, Stockholm SE-10691, Sweden}

\author{Boris Svistunov} 
\affiliation{Department of Physics, University of Massachusetts, Amherst, MA 01003, USA}
\affiliation{Wilczek Quantum Center, School of Physics and Astronomy and T. D. Lee Institute, Shanghai Jiao Tong University, Shanghai 200240, China}

\begin{abstract} 
While the properties of standard (single-component) superfluids are well understood, principal differences arise in a special type of multicomponent systems---the so-called Borromean supercounterfluids---in which (i) supertransport is possible only in the counterflow regime and (ii) there are three or more counterflowing components. Borromean supercounterfluids's correlation and topological properties distinguish them from their single- and two-component counterparts. 
 The component-symmetric case characterized by a distinctively different universality class of the supercounterfluid-to-normal phase transition is especially interesting. Using the recently introduced concept of compact-gauge invariance as the guiding principle, we develop the finite-temperature description of Borromean supercounterfluids in terms of an asymptotically exact long-wave effective action. We formulate and study Borromean XY and loop statistical models, capturing the universal long-range properties and allowing us to perform efficient worm algorithm simulations. Numeric results demonstrate perfect agreement with analytic predictions. Particularly instructive is the two-dimensional case, where the Borromean nature of the system is strongly manifested while allowing for an asymptotically exact analytic description.

\end{abstract}

\maketitle

\section{Introduction}
The phenomenon of counterflow superfluidity (a.k.a. supercounterfluidity)---taking place in a multicomponent system under the conditions when the net flow of atoms is either arrested by Mott physics \cite{kuklov2003counterflow} or becomes normal due to the finite-temperature proliferation of multicomponent (composite) vortices \cite{babaev2002phase}---has been of substantial interest for two decades \cite{kuklov2003counterflow,babaev2002phase,Duan2003,Altman2003,kuklov2004superfluid,Smiseth2005field,Dahl2008preemptive,Hubener2009,Powell2009,Hu2009,Herland2010phase,Menotti2010,Hu2011,Schachenmayer2015,soyler2009sign,Venegas_Gomez2020B,deParny2021,Basak2021} (for an introduction, see \cite{sbpbook}). The first experimental implementation of a supercounterfluid state---in a 6-site chain of two-component ultracold bosons---was reported recently \cite{Zheng2024SCF_exp}. 

The vast majority of previous theoretical work deals with the two-component case. Nevertheless, the general multicomponent case was not only known to exist in principle \cite{sbpbook} but was also implemented numerically (for the three-component Bose-Hubbard model) and dubbed ``Borromean" in Ref.~\cite{blomquist2021borromean}. Furthermore, it was also established \cite{Bojesen2013time,Bojesen2014phase} that a special (frustrating) type of intercomponent Josephson coupling modifies a three-component supercounterfluid into a very peculiar state---termed a Borromean metal at a finite temperature $T$ \cite{babaev2023hydrodynamics}---with a specific topological-like intercompoment ordering implying, in particular, broken time-reversal symmetry (BTRS). 
The recent experiments \cite{Grinenko2021state,shipulin2023calorimetric} reported observation of this state of matter in an iron-based material \bkfa, where the spontaneous magnetic fields arise due to persistent local counterflow currents. Another crucial aspect of such systems---the existence of elementary (a.k.a. fractional \cite{Babaev2002vortices}) and composite vortices (i.e., a bound state of several fractional vortices)---was also observed in \bkfa ~\cite{Iguchi2023}.
This state of matter comes with a large number of other unconventional properties that are currently under investigation.

 Despite impressive theoretical and experimental progress, a crucial question about Borromean counterfluids has not been addressed until very recently: What is the effective long-wave description of the system? In many respects, a two-component ($N = 2$) supercounterfluid maps onto a single-component superfluid. The Borromean case ($N \geq 3$) is very different. Think, for simplicity, about the Borromean ground state. 
Since the net matter flow is arrested, the number of independent phonon modes has to be equal to $N-1$, suggesting that the number of independent order parameters---defining the number of pairs of canonically conjugate fields in the Hamiltonian formalism---is also $N-1$. On the other hand, the number of elementary vortices, or, equivalently, elementary counterflow persistent-current states equals $N$, as if there were $N$ independent order parameters. In the most instructive case, when the system features permutation symmetry for all the $N\geq 3$ components, these $N$ elementary vortices are distinctively different and absolutely equivalent to each other in terms of their properties, including the energies.

In a recent work by some of us \cite{babaev2023hydrodynamics}, progress has been made in understanding the guiding principle, allowing the construction of an effective hydrodynamic description of Borromean ground states
by introducing the notion of special gauge redundancy of the theory [see Eq.~(\ref{gauge}) below], the so-called compact-gauge invariance. As opposed to standard local U(1) symmetry, it does not require the presence of an additional gauge field. Instead, it is a symmetry property of the effective long-wave Hamiltonian of the $N$-component system. By Noether's theorem, the compact-gauge symmetry guarantees the invariance of the net local density of the matter, thus enforcing the counterflow character of the system's dynamics.

In this paper, we show that the principle of compact-gauge invariance, formulated initially in the context of zero-temperature hydrodynamics, proves equally (if not more) crucial at a finite $T$. It readily allows us to formulate a finite-temperature description of Borromean supercounterfluids in terms of an asymptotically exact long-wave effective action. Guided by this principle, we construct generalized XY and loop (a.k.a. J-current) statistical models capturing the universal long-range properties of Borromean supercounterfluids. Analytic theory based on the effective long-range action demonstrates perfect agreement with the numerically exact results for a minimal loop model simulated by worm algorithm. Much of our analytic and numeric attention is paid to the two-dimensional (2D) case, where the Borromean nature of the system is strongly manifested in long-range off-diagonal correlations as well as in the equilibrium statistics of counterflow supercurrent states and where the Berezinskii--Kosterlitz--Thouless (BKT)--type criticality is driven by well-isolated topological defects thus allowing for an asymptotically exact analytic description. 

The rest of the paper is organized as follows. In Sec.~\ref{sec:long_wave}, we use the principle of compact-gauge invariance to formulate an effective description of an $N$-component counterflow superfluid.
In Sec.~\ref{subsec:B_XY}, we introduce the Borromean XY model, the minimal model capturing the corresponding universality class. In Sec.~\ref{subsec:bor_loop_model}, we establish the relationship between the XY-type and loop (a.k.s. J-current) Borromean models and representations. Here, the principle of compact-gauge invariance allows us to directly introduce Borromean loop models and, in particular, formulate the minimal one illustrated in Fig.~\ref{fig:bond_conf}. All our numeric data are obtained by worm-algorithm simulations of this minimal loop model.

In Sec.~\ref{sec:vortices}, we analyze topological structures---vortices and persistent currents---in Borromean models. In Sec.~\ref{sec:phase_twist_windings}, we study the response to the phase twist. We show (see Fig.~\ref{fig:C_err}) how the dependence of phase twist response on the aspect ratio can be used to probe the equilibrium statistics of supercurrent states.
In Sec.~\ref{sec:phase_field_fluct}, we study off-diagonal correlations---both analytically and numerically. Numeric data demonstrate perfect agreement with analytic predictions.

In Sec.~\ref{sec:BKT}, we develop the theory of BKT-type transition in a Borromean system, corroborating our results with numeric simulations. Here, we find that while renormalization of the superfluid stiffness by vortex-antivortex pairs has a Borromen character---reflecting Borromean properties of the underlying vortex plasma, the equations describing the renormalization flow can be formally cast into the standard Kosterlitz-Thouless form by absorbing the Borromean factor into the effective vortex pair concentration.

In Sec.~\ref{sec:discussion}, we discuss the obtained results and future research directions.

\section{Effective  description}
\label{sec:long_wave}

\subsection{Long-wave action}
\label{subsec:action}

The form of effective long-wave action for the finite-temperature $N$-component Borromean supercounterfluid can be readily established by utilizing formal correspondence with the $N$-component superfluid, where one degree of freedom is implicitly removed by compact-gauge redundancy \cite{babaev2023hydrodynamics}. The effective long-wave finite-temperature action for the $N$-component superfluid is a generalization of the single-component action. Here the effective degrees of freedom are $N$ fields of superfluid phases, $\theta_\alpha$, $\alpha=1,2,\ldots , N$. The [U(1)]$^N$ invariance of the theory requires that the action density be a function of gradients $\nabla \theta_\alpha$. At the same time, the long-wave regime means that we can confine ourselves with only leading---{\it i.e.}, quadratic---Taylor expansion of the action density in powers of $\nabla \theta_\alpha$. 
The key constraint distinguishing Borromean supercounterfluid from the $N$-component superfluid with a drag between components (caused by mixed gradient terms) is the requirement of invariance of the action with respect to the compact-gauge transformation
\be
\forall \alpha : \quad \theta_\alpha ({\bf r}) \, \to \,  \theta_\alpha ({\bf r}) \, +\, \varphi ({\bf r}) \, ,
\label{gauge}
\ee
where $\varphi ({\bf r})$ is an arbitrary field of phase defined modulo $2\pi$ and allowed to have tological defects (point vortices in 2D and vortex lines in 3D). 

The microscopic origin of the requirement (\ref{gauge}) can be rather different. If the system has the counterflow ground state so that the individual phases $\theta_\alpha$ cannot be unambiguously defined from the very outset, then Eq.~(\ref{gauge}) is merely a formal mathematical framework allowing one to treat on equal grounds all the $N$ components---as opposed to selecting one of them as the ``phase reference frame" and dealing with relative phases; see Eq.~(\ref{grtrans}) below. If the ground state of the system is the $N$-component superfluid and the supercounterfluid regime is enforced by the proliferation of composite vortices (having the same winding number $\pm 2\pi$ for each of $\theta_\alpha$), so that the phases $\theta_\alpha$ have at least some clear microscopic meaning. It now directly reflects the proliferation of composite vortices, meaning that they cost no energy at the macroscopic level.

Independently of its microscopic origin, Eq.~(\ref{gauge}) implies that the density of the long-wave action has the form 
\begin{equation}
{\cal A}  =  {1  \over 2 }  \sum_{\alpha <  \beta}  \, \Lambda_{\alpha \beta}\, (\nabla \theta_\alpha \! - \! \nabla\theta_\beta)^2   .
\label{A_general}
\end{equation}
The partition function describing the long-wave off-diagonal correlations in the system thus has the form
\begin{equation}
Z =\int   e^{-A[\{ \theta \}]} \, \prod_{\alpha=1}^{N}{\cal D} \theta_\alpha  \, , \quad A[\{ \theta \}]  = \int {\cal A}\, d^d r\, .
\label{partit_B}
\end{equation}
Here $\{ \theta \} \equiv (\theta_1, \theta_2, \ldots \theta_N)$

In what follows, we will be dealing with the component symmetric case 
\begin{equation}
{\cal A}  =  {\Lambda  \over 2 }  \sum_{\alpha < \beta} \, (\nabla \theta_\alpha \! - \! \nabla\theta_\beta)^2   \,  .
\end{equation}
The symmetry between components guarantees that the phase transition from the counterfluid to normal fluid happens for all the components simultaneously.
For future convenience of tracing correspondence with the single-component case, we will write the action as
\begin{equation}
{\cal A}  =  {\Lambda_s  \over 2(N-1) }  \sum_{\alpha < \beta} \, (\nabla \theta_\alpha \! - \! \nabla\theta_\beta)^2   \,  .
\label{A_sym}
\end{equation}
where
\be
\Lambda_s = \Lambda(N-1)
\label{eq:Lambda_s}
\ee
We will see that, in many respects, the parameter $\Lambda_s$ will play a role very similar to that of the superfluid stiffness of a single-component system.

Even in the component-symmetric case (\ref{A_sym}), democratically treating all the components at the expense of gauge-redundancy is not always convenient.
The redundancy can be removed by working with gauge-invariant quantities---relative phases of $(N-1)$ components with respect to the phase of the $N$-th one \footnote{Note that Eq.~(\ref{grtrans}) is consistent with the requirement that any redefinition of phases should respect the $2\pi$-periodicity of the original phases.}:
\be
\phi_\alpha \, =\, \theta_\alpha - \theta_N \qquad \quad  [\, \alpha = 1, 2,3, \ldots, (N-1) \, ] \, .
\label{grtrans}
\ee
In terms of these relative phases, we get (the  integration over $\theta_N$ factors out and is omitted in what follows since it corresponds to nothing but gauge redundancy)
\begin{equation}
\tilde{Z} =\int   e^{-\tilde{A}[\{ \phi \}] } \, \prod_{\alpha=1}^{N-1}{\cal D} \phi_\alpha  ,  \quad \tilde{A} [\{ \phi \}]  = \int \tilde{\cal A}\, d^d r\, , \label{tilde_partit_B}
\end{equation}
\be
\begin{aligned}
 &\tilde{\cal A} =  {\Lambda _s \over 2(N-1) }  \sum_{\alpha=1}^{N-1}  (\nabla \phi_\alpha )^2 \\
 &\,+\, {\Lambda_s  \over 2(N-1) }  \sum_{\alpha < \beta < N}   (\nabla \phi_\alpha \! - \! \nabla\phi_\beta)^2 \, .
\end{aligned}
\ee
It is good to observe that we could arrive at the very same parameterization by utilizing gauge redundancy---working in the gauge where $\theta_N\equiv 0$ and thus $\phi_\alpha \equiv \theta_\alpha$.

 For our future purposes, it is convenient to identically rewrite the expression for $ \tilde{\cal A} $ as
 \be
  \tilde{\cal A} =  {\Lambda_s  \over 2 }  \sum_{\alpha=1}^{N-1}  (\nabla \phi_\alpha )^2 \,-\, {\Lambda_s \over (N\! -\! 1)} \! \sum_{\alpha < \beta < N} \!  \nabla \phi_\alpha \! \cdot \! \nabla\phi_\beta \, .
\label{tilde_A_B}
\ee 
\

In such a parameterization, the model has a form similar to that of $(N-1)$-component superfluid with specially fine-tuned drag couplings to guarantee that   $(N-1)$-component composite vortex having the same winding number in each of the phases would have the same energy as the elementary vortex having the winding number  in the phase of only one of the $N-1$ components. We should realize, however, that such fine-tuning is akin to accidental degeneracy rather than symmetry: It cannot be achieved at the level of certain fixed parameters of the system because changing the temperature would detune the system from this special regime. The only option is to move along a special line $p=p(T)$ in the $(p,T)$ phase diagram of the system, with parameter $p$ allowing to re-adjust the drag to achieve the correspondence with the $N$-component Borromean system.

In the Borromean supercounterfluid, the fields of relative phases play a key role in defining the concept of genuine (in 3D) or algebraic (in 2D) off-diagonal long-range order. The simplest off-diagonal correlator---to be referred to as composite density matrix---is defined as
\begin{equation}
\rho ({\bf r}) \,  = \, \langle e^{i \phi_\alpha ({\bf r}) - i\phi_\alpha
(0)} \rangle \, .
\label{rho_Bor}
\end{equation}

\subsection{Borromean XY model}
\label{subsec:B_XY}

The long-wave effective action is perfect for analytic treatments. For efficient numeric simulations, one would prefer to deal with minimal discrete models that capture the same long-wave behavior while being particularly suited for advanced algorithms. In the single-component case, the XY model and its J-current counterparts are the models of choice.

 Let us now generalize the lattice XY model to the lattice model of Borromean counterfluid. 
The guiding principle is compact-gauge invariance as with the long-wave effective action (\ref{A_general}). Hence, we formulate   the following expression for the Hamiltonian of the Borromean XY model:
\be
    H\, =\, - \sum_{\alpha < \beta}\sum_{\langle s_1 s_2 \rangle} 
    J_{\alpha\beta}\cos{(\theta_{\alpha s_1} - \theta_{\alpha s_2} - \theta_{\beta s_1} + \theta_{\beta s_2})} \, .
\label{H_Bor}
\ee
The key ingredient here is the compactness of the individual phase fields $\theta_{\alpha s}\in [0, 2\pi]$; $\alpha$ is the component index on the site $s$ of the hypercubic lattice in $d$ dimensions. 
The symbol $\langle \ldots \rangle$ restricts summation over the pairs of sites to the pairs of nearest neighbors.

The lattice analog of the compact-gauge transformation (\ref{gauge}) is
\be
\forall s, \, \forall \alpha : \quad \theta_{\alpha s} \, \to \,  \theta_{\alpha s}  \, +\, \varphi_s \, ,
\label{gauge_discr}
\ee
where $\varphi_s$ is an arbitrary real number. Hamiltonian (\ref{H_Bor}) is invariant with respect to his transformation.

In what follows, we will be interested only in the case when all the couplings are the same ($J_{\alpha\beta} = J_s/(N\! -\! 1) > 0$):
\be
H\, =\, -\frac{J_s}{2(N\! -\! 1)}\sum_{\alpha \neq \beta}\sum_{\langle s_1 s_2 \rangle} 
    \cos{(\theta_{\alpha s_1} - \theta_{\alpha s_2} - \theta_{\beta s_1} + \theta_{\beta s_2})} . 
\label{H_sym}
\ee

In the direct analogy with Eq.~(\ref{grtrans}), the gauge freedom can be removed by working with relative phases (note that while changing variable notations, one should retain the $2\pi $ periodicity of the original phases):
\be
\phi_{\alpha s} \, =\, \theta_{\alpha s} - \theta_{N s} \qquad \quad  [\, \alpha = 1, 2,3, \ldots, (N-1) \, ] \, .
\label{grtrans_latt}
\ee
In this representation, the Hamiltonian (\ref{H_sym}) becomes
\begin{align}
    &H = -\frac{J_s}{(N\! -\! 1)}\sum_{\alpha}^{N - 1}\sum_{\langle s_1 s_2 \rangle} 
    \cos{(\phi_{\alpha s_1} - \phi_{\alpha s_2})} \nonumber \\
    &
     -\frac{J_s}{2(N\! -\! 1)}\sum_{\alpha \neq \beta}^{N - 1}\sum_{\langle s_1 s_2 \rangle} 
   \cos{(\phi_{\alpha s_1} - \phi_{\alpha s_2} - \phi_{\beta s_1} + \phi_{\beta s_2})} .
\end{align}

The lattice composite density matrix---lattice counterpart of (\ref{rho_Bor})---is
\begin{equation}
\rho_{s_1 s_2} \,  = \, \langle e^{i \phi_{\alpha s_1} - i \phi_{\alpha s_2} } \rangle \, .
\label{rho_Bor_latt}
\end{equation}

\subsection{Borromean loop models/representations} 
\label{subsec:bor_loop_model}

In the theory of superfluidity, of particular importance are loop (a.k.a. J-current) models \cite{sbpbook}. The loop models emerge (or can be viewed as) dual representations of certain lattice models written initially in terms of the phases $\theta_{\alpha j}$. Equivalently, a loop model can, in principle, be converted into an equivalent $\theta$-model. 

Loop models/representations provide an alternative perspective on the nature and mechanism of superfluid ordering and serve as the basis for most efficient numeric simulations with the worm algorithm \cite{Prokofev1998a,Prokofev1998b,Prokofev2001}. All these aspects are essential in the Borromean case as well. Furthermore, there is one additional circumstance rendering loop representations particularly relevant. In the $\theta$-representation, the visuality of component symmetry comes at a considerable price of gauge redundancy. As we will see in what follows, loop representation is free of gauge redundancy while being explicitly component-symmetric. The counterfluid regime is enforced by the most visual constraint requiring that (for each bond)  the net bond current be zero.

\begin{figure}[t]
\includegraphics[width=0.7 \columnwidth]{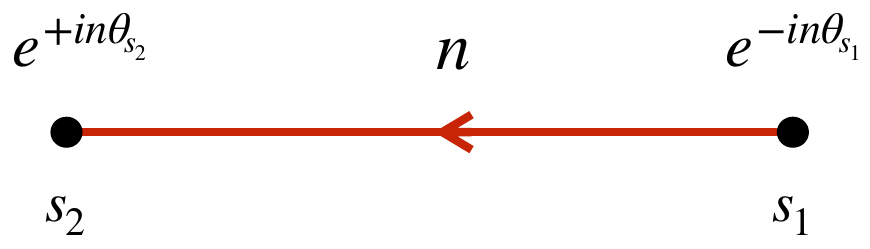}
\caption{Bond current notation. The current of a given color and finite integer strength $n>0$ goes from the site $s_1$ to the site $s_2$. (Color subscripts are suppressed for clarity.)}
\label{Bond_current}
\end{figure}

\begin{figure}[t]
        \vspace{0.2cm}
\includegraphics[width=0.7\columnwidth]{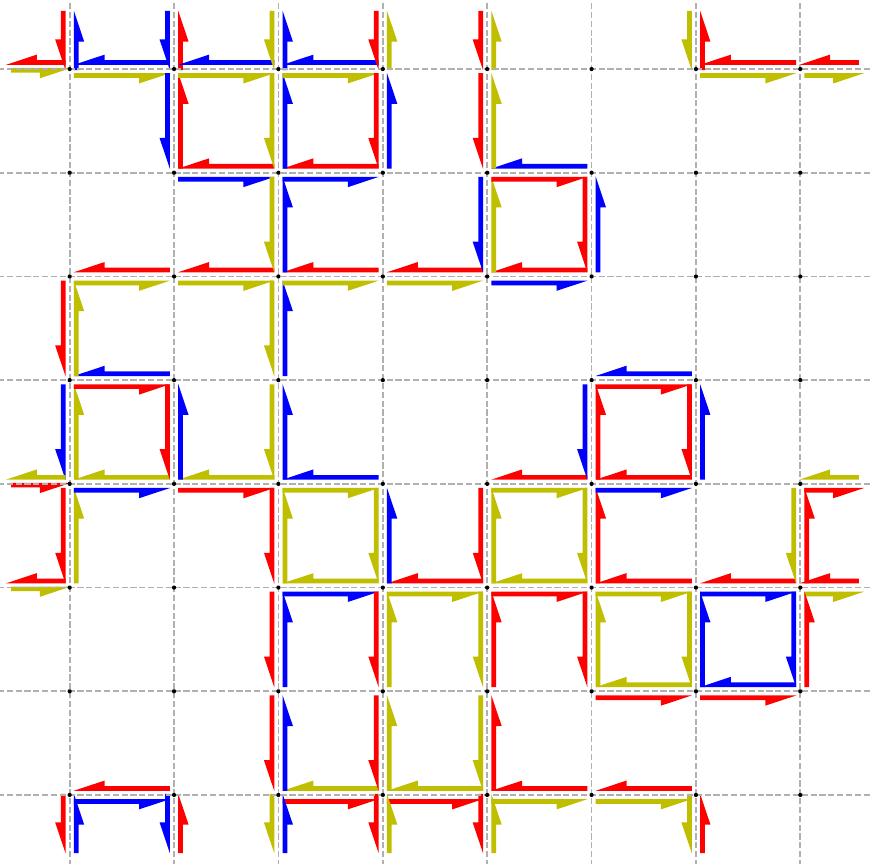}\\ 
    \vspace{0.6cm}
\includegraphics[width=0.7\columnwidth]{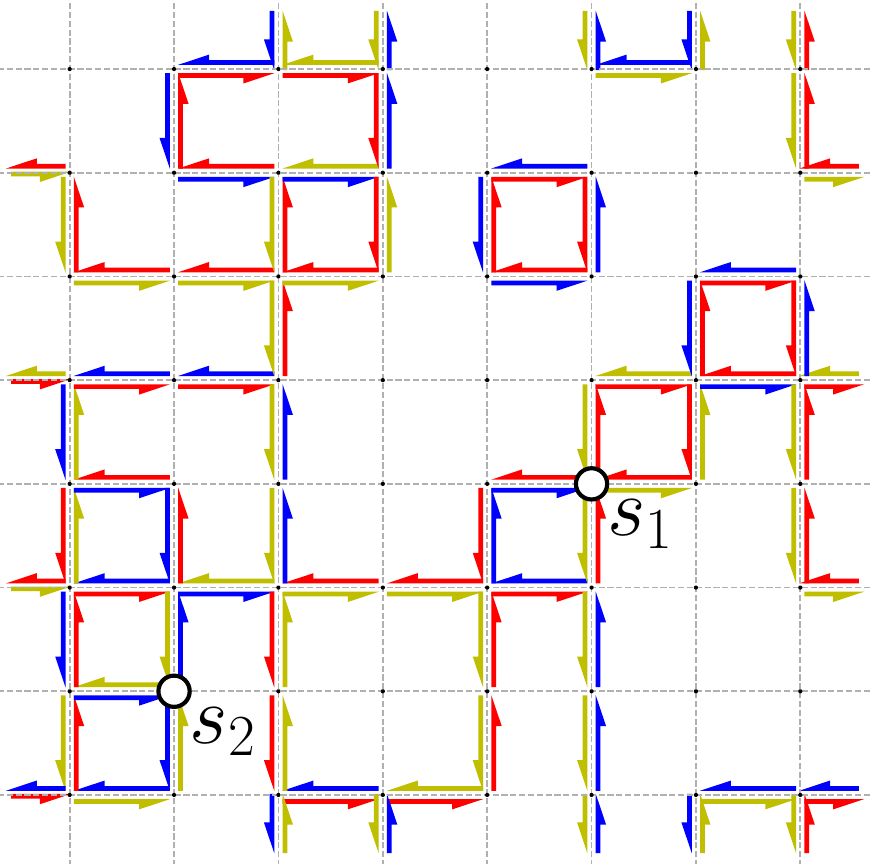}
    \vspace{0.25cm}
    \caption{Minimal Borromean loop model: Typical bond configurations for the 2D three-component system of the linear size $L=8$ with periodic boundary conditions. Top: Bond configuration with closed loops, contributing to the partition function (\ref{J_current}) with the constraint (\ref{gauuge_inv_loop}). Bottom: Worm diagram with head and tail (white circles), contributing to the composite density matrix (\ref{rho_Bor_latt}). Each bond is either empty [contributing the trivial factor of $F=1$ to the partition function (\ref{J_current})] or contains two counter-currents, in which case it contributes the factor $F=P$, where $P\in (0,1) $ is a certain constant---the only control parameter of the model. At each site, apart from the two worms, $s_1$ and $s_2$, the algebraic sum of currents of each component is zero.
    Having at least three components is crucial for the configurational space to principally differ from the minimal {\it single-component} model. With only two counterflowing components, the zero-current constraint enforces exact equivalence with the single-component configurational space.}  
    \label{fig:bond_conf}
\end{figure}

A standard way of arriving at loop representation is by doing high-temperature expansion of the partition function of a certain XY-type model. We will follow this route with an observation that an explicit formulation of the model in terms of the Hamiltonian is unnecessary. All we need are some general properties of XY-type models (be those supercounterfluids or $N$-component superfluids) implying the following structure of the partition function:
\be
Z\, =\, \int \left[ \prod_b Q\left( \{ \nabla \theta\}_b \right)\right]  \prod_{\alpha j} d\theta_{\alpha j} \, .
\label{Z_gen_bond}
\ee
Here, the subscript $b$ runs over all the bonds of the lattice, and the symbol $ \{ \nabla \theta\}_b$ denotes the set of all lattice gradients of the lattice fields $\theta_{\alpha j}$ associated with this bond:
\be
\{ \nabla \theta \}_b \, =\, \left( \nabla \theta_1^{(b)}, \nabla \theta_2^{(b)}, \ldots ,\nabla \theta_N^{(b)} \right) \, ,
\label{lat_grad_set}
\ee
\be
\nabla \theta_\alpha^{(b)} \, =\, \theta_{\alpha s_2} - \theta_{\alpha s_1} \ ,
\label{lat_grad}
\ee
with $s_1\equiv s_1^{(b)}$ and $s_2\equiv s_2^{(b)}$ labeling two lattice sites connected by the bond $b$. The choice of which of the two is $s_1$ and $s_2$, respectively, is a matter of convention; it plays no role in further discussion because the factor $Q$ is supposed to be an even function of $\nabla \theta_\alpha^{(b)}$.

In direct correspondence with the continuous-space case, the dependence on the gradients guarantees global U(1)$^N$ symmetry of the theory---defining feature of the $N$-component superfluid system. The lattice-specific property coming from the compactness of the site variables, $\theta_{\alpha j}\in [0, 2\pi]$, is the necessity for the factor $Q$ to be a $2\pi$-periodic function of each $\theta_{\alpha j}$ and thus of each $\nabla \theta_\alpha^{(b)}$. 

Because of this periodicity, we can expand $Q\left( \{ \nabla \theta\}_b \right)$ into the Fourier series (for the sake of clarity, we suppress the sub- and super-scripts $b$)
\be
Q\left( \{ \nabla \theta\} \right) \, =\,  \sum_{\bf n} F_{\bf n} \exp \left( i \sum_{\alpha = 1}^N n_\alpha \nabla \theta_\alpha \right)  \, .
\label{Q_Fourier}
\ee
Here 
\be
{\bf n} \, =\, (n_1, n_2, \ldots, n_N)
\label{set_n}
\ee
is an $N$-component integer vector labeling Fourier coefficients $F_{\bf n}$.

Now we improve notation that will be important for the next step. From now on, all our integers $n_\alpha$ will be non-negative. In contrast, the possibility of having two different signs for any nonzero $n_\alpha$ will be accounted for by a binary subscript that is conveniently represented graphically as an arrow (of the ``color" $\alpha$) attached to the bond and pointing from the vertex $s_1$ associated with the exponential $e^{-in_\alpha \theta_{\alpha s_1}}$ to the vertex $s_2$ associated with the exponential $e^{+in_\alpha \theta_{\alpha s_2}}$. Correspondingly, we will be saying that there is an integer J-current going from the vertex $s_1$  to the vertex $s_2$; for an illustration, see Fig.~\ref{Bond_current}. In the $n=0$ case, we will be saying that the bond current is absent and, correspondingly, will not be introducing any graphical elements. In the context of this notation, we introduce the variable $j_\alpha^{(b)}$ standing for the current of the component $\alpha$ on the bond $b$ (as well as its absence), in accordance with the convention of Fig.~\ref{Bond_current}. 

We then introduce the notion of the divergence of the bond currents (of a given color $\alpha$) at the site $s$, denoting it with the symbol $\nabla \cdot {\bf j}_{\alpha s}$, which corresponds to the algebraic---outgoing minus incoming---sum of all the bond currents of the given color associated with the given site. The importance of the notion becomes immediately clear once the Fourier expansion (\ref{Q_Fourier}) is substituted into (\ref{Z_gen_bond}), allowing one to explicitly perform the integrals over all angles. The integration over $\theta_{\alpha s}$ produces exact zero unless $\nabla \cdot {\bf j}_{\alpha s}$, in which case the integration produces the factor $(2\pi)^N$; this global factor plays no statistical role and will be omitted in what follows.

This way we arrive at the loop (a.k.a. J-current) representation of the partition function:
\be
Z\, =\, \sum_{\{ j \}}^{\nabla \cdot {\bf j} =0} \prod_b F(\{ j\}_b) \, .
\label{J_current}
\ee
Here, the sum is over all the configurations of bond currents satisfying the zero-divergence condition at each site and for each component:
\be
\forall s , ~\forall \alpha: \quad \nabla \cdot {\bf j}_{\alpha s}\, =\, 0\, .
\label{div_zero}
\ee
The factor $F(\{ j\}_b)$ is the Fourier coefficient of Eq.~(\ref{Q_Fourier}) parameterized in accordance with convention of Fig.~\ref{Bond_current}. 

So far, our analysis did not distinguish between the $N$-component superfluid and the supercounterfluid. To make sure we are dealing with the latter rather than the former, we need to impose the condition of the compact-gauge invariance on the bond factors $Q\left( \{ \nabla \theta\}_b \right)$ in the partition function (\ref{Z_gen_bond}). 

In terms of the Fourier expansion (\ref{Q_Fourier}), compact-gauge transformation on one of the two sites associated with the bond $b$ leads to 
\be
F(\{ j\}_b)\, \to\, F(\{ j\}_b) \exp \left( i \varphi \sum_{\alpha = 1}^N j_\alpha^{(b)} \right) \, .
\label{gauge_Fourier}
\ee
Hence, the compact-gauge invariance  implies the condition
\be
\forall b: \quad \sum_{\alpha = 1}^N \, j_\alpha^{(b)} \, =\, 0 \, .
\label{gauuge_inv_loop}
\ee
This condition---requiring that the algebraic sum of all the currents on each bond be always zero---is very visual.

Now we can formulate the minimal loop Borromean model. The minimal number of components is three. The minimal number of nonzero bond currents necessary to satisfy the constraint (\ref{gauuge_inv_loop}) is two. The minimal nonzero magnitude of the bond current is one.
Each empty bond contributes the trivial factor of $F=1$ to the partition function (\ref{J_current}).
Each bond with two counter-currents contributes the factor $F=P$, where $P\in (0,1) $ is a certain constant---the only control parameter of the model.
This brings us to the configuration space shown in Fig.~\ref{fig:bond_conf}. It is clearly seen that having at least three different components is crucial for the configurational space to be principally different from that of the minimal {\it single-component} model. In the case of only two counterflowing components, the zero-current constraint, requiring that the magnitudes of the two bond currents be precisely equal, enforcing mapping between the two- and single-component configuration spaces.

\subsection{Numeric protocol}
\label{sunsec:Numeric_protocol}

All our numeric data are obtained by simulating the multicomponent two-dimensional loop model illustrated in Fig.~\ref{fig:bond_conf}. We use the standard worm algorithm for classical loop models \cite{Prokofev2001}, which gives us a natural access to the composite density matrix (\ref{rho_Bor_latt}) via the statistics of the two worms, $s_1$ and $s_2$ (see bottom panel in Fig.~\ref{fig:bond_conf}). Worm algorithm also provides us with simple access to the superfluid stiffness $\Lambda_s$ via the statistics of loop windings; see Sec.~\ref{subsec:Pollock_Ceperley} for the details. To sample a system of linear size $L \sim 10^2$ with small statistical error bars (see numeric results throughout the paper), we perform $\sim 10^{13}$ Monte Carlo steps upon thermalization.

\section{Vortices and persistent currents}
\label{sec:vortices}

\subsection{Extremal solutions and topological charges} \label{subsec:topological_charges}

Up to the compact-gauge freedom, the extremal configurations of the fields---nullifying the first variational derivatives of the action $A[\{ \theta \}]$---satisfy the set of equations
\be
\forall \alpha:  \quad \Delta \theta_\alpha^{\rm (top)} \, =\, 0 \, .
\label{extremal_theta}
\ee
The same statement can be made in the gauge-invariant form in terms of the relative phases and the action $\tilde{A}[\{ \phi \}]$:
\be
\forall \alpha:  \quad \Delta \phi_\alpha^{\rm (top)} \, =\, 0 \, .
\label{extremal_phi}
\ee
Because of the compactness of the fields of the phase, Eqs.~(\ref{extremal_theta}) and (\ref{extremal_phi}) can have topologically nontrivial solutions, provided the system is multiply connected or contains a topological defect (vortex). In the latter case, Eqs.~(\ref{extremal_theta}) and (\ref{extremal_phi}) take place away from the vortex core. In the former case, topologically nontrivial solutions correspond to persistent currents. In both cases, any topologically nontrivial extremal solutions can be parameterized as
\be
\theta_\alpha^{\rm (top)} ({\bf r}) \, =\, M_\alpha \varphi_{\rm top}({\bf r}) \, , 
\label{top_par}
\ee
where $M_\alpha$ are certain integer numbers and the field $\varphi_{\rm top}({\bf r})$ satisfies the Laplace equation
\be
\Delta \varphi_{\rm top} \, =\, 0
\label{varphi_lap}
\ee
and the condition
\be
\oint_{C}d{\bf l} \cdot  \nabla \varphi_{\rm top}  \, =\, 2\pi \, .
\label{varphi_wind}
\ee
with respect to a certain contour $C$.

The integer vector
\be
{\bf M} \, =\, (M_1,M_2, \ldots , M_N)  
\label{vector_m}
\ee
characterizes the topological charge of the vortex/persistent current. By the compact-gauge invariance, vector ${\bf M}$ is defined modulo vector $\hat{1}$ consisting of unities:
\be
\hat{1} \, =\, (1,1, \ldots , 1) \, . 
\label{hat_1}
\ee
Because of the linearity of Eqs.~(\ref{extremal_phi})--(\ref{varphi_wind}), we can extend the notions of the topological charge and corresponding extremal solutions to a group of more than one vortex, provided the group is localized in space. We are interested in distances much larger than the linear size of the group. At such distances, the group of vortices is indistinguishable from a single vortex of the charge equal to the sum of all the individual charges modulo $\hat{1}$. We thus can speak of the modular arithmetic of topological charges.

For the sake of unambiguously labeling all the topological charges, we can also introduce gauge-invariant relative topological charges 
\be
\tilde{M}_\alpha = M_\alpha - M_N \, , \qquad (\alpha < N)\,  ,
\label{M_rel}
\ee
combining them into the $(N-1)$-component integer vector $\tilde{\bf M}$.

\subsection{Energetics and statistics}

The extremal character of the vortex/persistent-current solutions, in combination with the bilinearity of the action, allows one to decompose the fields $\theta_\alpha$ into the {\it statistically independent} topological, $\theta_\alpha^{\rm (top)}$, and regular, $\theta_\alpha^{(0)}$, parts:
\be
\theta_\alpha \, =\, \theta_\alpha^{\rm (top)} + \theta_\alpha^{(0)} \, ,
\label{theta_decomp}
\ee 
\be
A \, =\, A_{\rm top}\, +\, A[\{ \theta^{(0)} \}] \, ,
\label{A_decomp}
\ee
\be 
A_{\rm top}({\bf M}) \, =\, A[\{ \theta^{\rm (top)} \}] \, =\, 
 {\Sigma({\bf M})\,  \Lambda_s \over 2} \int (\nabla \varphi_{\rm top})^2 \, d^d r\, ,
\label{A_top}
\ee
where 

\be
\Sigma({\bf M})\, = \, {1\over N-1} \sum_{\alpha < \beta} \, (M_\alpha-M_\beta)^2 \, ,
\label{Sigma_M}
\ee

The idea behind including the pre-factor $(N-1)^{-1}$ into (\ref{Sigma_M}) is that we want to normalize $\Sigma({\bf M})$ in such a way that it equals unity for elementary vortices/persistent-current states having topological charge $|{\bf M}|=1$.   Correspondingly, $\Lambda_s$ (\ref{eq:Lambda_s}) plays the role of superfluid stiffness characterizing energetics of elementary vortices.

Sometimes it is also useful to work with a gauge-invariant expression for the factor $\Sigma$ in terms of gauge-invariant charges $\tilde{M}_\alpha$ (\ref{M_rel}):
\be
\Sigma({\bf M})\, = \, \tilde{\Sigma}(\tilde{\bf M}) = \sum_{\alpha=1}^{N-1} \tilde{M}_\alpha^2\!  -
{2\over N\! -\! 1} \!\! \sum_{\alpha<\beta<N} \!\tilde{M}_\alpha \tilde{M}_\beta \, .
\label{Sigma_rel}
\ee
While having the same values for corresponding values of their arguments, the functions $\Sigma$ and $\tilde{\Sigma}$ differ by their functional form and therefore are denoted with different symbols.

\subsection{Isolated vortex. Nelson-Kosterlitz relation}
\label{subsec:isolated_v}

From equation (\ref{A_top}) we see that the leading term for the ``energy"  of an isolated vortex  is given by (per unit length of a straight vortex line in 3D)
\be
E_{\rm vort} = \pi \Sigma({\bf M})\,  \Lambda_s \ln (L/ l_0) \, ,
\label{v_vort}
\ee
where $L$ is a typical linear system size and $l_0$ is the characteristic microscopic cutoff. Here and in what follows, the term ``energy" in the context of vortices and supercurrent states is used as a shorthand for ``free energy measured in units of temperature."

For an elementary vortex, $\Sigma=1$, the expression is formally the same as in a single-component case, which, as we already mentioned, is merely the idea behind our definition of $\Lambda_s$, Eq.~(\ref{eq:Lambda_s}).
For a given (nonzero) topological charge, elementary vortices have the minimal possible energy and thus are energetically protected from decay/splitting into other vortices.

Because of entropic contributions in 2D, at  a certain critical value of $\Lambda_s$,  given by the counterpart of the universal Nelson-Kosterlitz relation \be
\Lambda_s^{\rm (BKT)} \, =\, {2\over \pi}\, ,
\label{NK_relation}
\ee
the
vortex proliferation transition takes place, and the state becomes normal.

\subsection{Energy of a vortex cluster} \label{sec:vortex_cluster}

Here, we confine ourselves to a 2D case (or, equivalently, a 3D system of straight parallel vortex lines), where generalization of the results of Sec.~\ref{subsec:isolated_v} is particularly straightforward.  The bilinearity of the effective action guarantees that the decomposition (\ref{theta_decomp}) into $\theta_\alpha^{\rm (top)}$ responsible for all the topological charges while satisfying Eq.~(\ref{extremal_theta}) and the regular part $\theta_\alpha^{(0)}$ works: The distributions of defects and $\theta_\alpha^{(0)}$ prove to be independent.  Furthermore, the linearity of Eq.~(\ref{extremal_theta}) implies a very simple structure of the solution---the superposition of solutions for individual vortices considered in Sec.~\ref{subsec:isolated_v}:
\be
\theta_\alpha^{\rm (top)} ({\bf r}) \, =\, \sum_{j} \, M_{\alpha j} \, \varphi_{\rm top}^{(j)}({\bf r}) \, . 
\label{top_par_j}
\ee
Here $j$ labels the individual vortices, $M_{\alpha j}$ is the topological charge associated with component $\alpha$ in the $j$-th vortex, and 
\be
\Delta \varphi_{\rm top}^{(j)} \, =\, 0\,,  \qquad 
\oint_{C_j}d{\bf l} \cdot  \nabla \varphi_{\rm top}^{(j)}  \, =\, 2\pi \, ,
\label{varphi_lap_j}
\ee
for any contour ${C_j}$ such that the $j$-th vortex is inside while all other vortices are outside it. Substituting (\ref{top_par_j}) into $A_{\rm top} \equiv A[\{ \theta^{\rm (top)} \}]$ and performing spatial integration results in the expression for the energy of the vortex cluster. The energy comes as a sum of pairwise contributions because of the bilinearity of the functional $A$.

Ignoring the overall charge term given by Eq.~(\ref{v_vort}), the interaction energy between two vortices, $i$ and $j$, of charges $M_{\alpha i}, M_{\alpha j}$ in the same component is given by
\be
E_{\text{pair,}\alpha\alpha} \, =\,  -2\pi\Lambda_sM_{\alpha i}M_{\alpha j}\ln{R_{ij}} \, +\,  \text{const}, \label{eq:pair_energy}
\ee
where $R_{ij}$ is the distance between the two vortices.
For two vortices with charges $M_{\alpha i}, M_{\beta j}$ in different components the same quantity is given by
\be
E_{\text{pair,}\alpha\neq \beta} \, =\,  2\pi\frac{\Lambda_s}{N-1} M_{\alpha i}M_{\beta j}\ln{R_{ij}} \, +\,  \text{const}. \label{eq:diff_pair_energy}
\ee
While intracomponent interactions are repulsive the intercomponent interactions are attractive,
although with a strength weaker by a factor $1/(N-1)$. Similar to the single-component case, the interaction has the form of a 2D Coulomb force, and the system can be thought of as a 2D plasma. This will, however, be a multicomponent  plasma (i.e. with several ``colors" of electric charges), which behaves very differently from the standard single-component one.

\subsection{Equilibrium statistics of supercurrent states}
\label{subsec:eq_supercurrents}

In certain cases, the supercurrent states can provide a noticeable or significant contribution to equilibrium statistics in a system in the form of torus/annulus or in a simulation box with periodic boundary conditions.

For simplicity, and with our future computational purposes in mind, here we will consider the case of a rectangular system with periodic boundary conditions. In the rectangular geometry, supercurrent states along different Cartesian directions are statistically independent of the regular fields $\theta_\alpha^{(0)}$ and each other. Therefore, it is sufficient to consider the statistics of supercurrent states along the $x$-axis. Here we have $\nabla \varphi_{\rm top}=2\pi/L_x$, with $L_x$ the  size of the system in $x$-direction, and from Eq.~(\ref{A_top}) we find the (non-normalized) statistical weight of the supercurrent 
with the charge ${\bf \tilde{M}}$ (we use gauge-invariant nomenclature to avoid double counting of the same states, which will be particularly important below):
\be
w_{\tilde {\bf M}} \, =\, e^{-2\pi^2 \tilde{\Sigma}(\tilde{\bf {M}})\Lambda_s V/L_x^2 } \, ,
\label{w_M}
\ee
where $V$ is the system volume. In a 3D system, the factor $V/L_x^2$ is macroscopical, so it makes sense to talk of the equilibrium supercurrents only if the transverse system sizes, $L_y$ and $L_z$, are very moderate and/or the size $L_x$ is anomalously large.

In 2D, we have 
\be
w_{\tilde {\bf M}}\, =\, e^{-2\pi^2 \tilde{\Sigma} ({\tilde {\bf M}})\Lambda_s (L_y/L_x) } \qquad (d=2) \, .
\label{w_M_2D}
\ee
In particular, at the  BKT-like transition point, using (\ref{NK_relation}), we find
\be
w_{\tilde {\bf M}}^{\rm (BKT)}\, =\, e^{-4\pi \tilde{\Sigma}(\tilde{\bf M}) (L_y/L_x) }  \, .
\label{w_M_2D_crit}
\ee
For a square system, 
\be
{w_{\tilde {\bf M}}\over w_{\tilde {\bf M} =0}} \, \leq \, e^{-4\pi} \ll 1 \qquad \mbox{(square system)}\, , 
\label{square_system}
\ee
and contributions from supercurrent states are tiny. However, working with aspect ratios $L_x/L_y \gg 1$, we can get the regime when $w_{\bf M} \sim 1$, which will allow us to probe the function $\Sigma({\bf M})$ via Eq.~(\ref{w_M_2D_crit}) and thereby observe quantitative predictions of our theory.

\section{Response to the phase twist and loop windings}
\label{sec:phase_twist_windings}

The value of the superfluid stiffness $\Lambda_s$, including its dependence on the system size, which is very important for tracing the critical behavior in the vicinity of the BKT transition---can be extracted from the system's response to the infinitesimal phase twist. In the loop/worldline representation, the response to the phase twist is directly related to the variance of the loop winding numbers, as was first observed by Pollock and Ceperley for path integral representation of bosons \cite{Pollock_Ceperley_relation}. Below we derive the Pollock-Ceperley relation for Borromean supercounterfluids. Meanwhile, we recall that the response to the 
twist phase is particularly informative in two dimensions \cite{PS_2000_two_definitions} (see also \cite{Melko2004_two_definitions,sbpbook}), where---using system's aspect ratio as a control parameter---we will employ it for detailed probing (the Borromean features of) the statistics of supercurrent states. 

\subsection{Response to the phase twist}
\label{subsec:resp_phase_twist}

For our purposes, it is sufficient to consider rectangular geometry with periodic boundary conditions and the phase twist, $\varphi_0$, applied only to one component and in one Cartesian direction. For definiteness, we select the $x$-direction and $N$-th component. The phase twist is a topological property, and the natural way to introduce it is by adding an extra term to $\theta_N^{\rm (top)}$
\[
\theta_N^{\rm (top)} \, \to \, \theta_N^{\rm (top)} + \varphi_0 {x\over L_x} \, ,
\]
in which case, integrating out the statistically independent fluctuations of the regular fields $\theta_\alpha^{(0)}$ and summing over the statistically independent supercurrent states along other Cartesian directions, we get the $\varphi_0$-dependent part of the partition function (up to an irrelevant factor):
\be
Z \, \propto \, \sum_{\bf \tilde{M}} w_{\bf \tilde{M}}(\xi)   \, , 
\qquad \xi = {\varphi_0 \over 2\pi} \, ,
\label{Z_xi}
\ee
where $w_{\bf \tilde{M}}(\xi)$ is obtained from $w_{\bf \tilde{M}}$ of Eq.~(\ref{w_M}) by replacing $\tilde{\Sigma} ({\tilde {\bf M}}) \, \to \,  \tilde{\Sigma} ({\tilde {\bf M}}, \xi)$, where 
\be
 \tilde{\Sigma} ({\tilde {\bf M}}, \xi) \, =\, \tilde{\Sigma} ({\tilde {\bf M}}) \, +\, 
{1\over N-1}\sum_{\alpha = 1}^{N-1} (\tilde{M}_\alpha - \xi)^2 \, .
\label{Sigma_xi}
\ee

In the theory of superfluidity [more generally, in any theory featuring topologically or genuinely broken global U(1) symmetry, an important role is played by the response function
\be
\Lambda^{(\varphi)}_x \, =\, - {L_x^2 \over V } \left. {\partial^2 \ln Z \over \partial \varphi_0^2} \right\vert_{\varphi_0=0} \, =\, - {L_x^2 \over V Z } \left. {\partial^2 Z \over \partial \varphi_0^2} \right\vert_{\varphi_0=0}   .
\label{Lambda_phi_def} 
\ee
The phase-twist response $\Lambda^{(\varphi)}_x$ is closely---but not always trivially---related to the superfluid stiffness $\Lambda_s$.
The same is true for the Borromean case. 
Doubly differentiating the partition function $Z$ with respect to the phase twist $\xi$ at then setting $\xi=0$, we obtain 
\be
\Lambda^{(\varphi)}_x \, =\, \Lambda_s  - \frac{4\pi^2\Lambda_s^2\,  V}{(N - 1)^2L_x^2} \left\langle \, \left(\sum_{\alpha = 1}^{N-1}\tilde{M}_{\alpha}\right)^2\right\rangle \, ,
\label{Lambda_phi}   
\ee
where the averaging is over all  configurations of the supercurrents:
\be
\langle \,  ( \ldots ) \, \rangle \, =\, 
{\sum_{\bf \tilde{M}} \, (\ldots) \, w_{\bf \tilde{M}} \over \sum_{\bf \tilde{M}}\,  w_{\bf \tilde{M}}} \, .
\label{av_def}
\ee
From the analysis of Sec.~\ref{subsec:eq_supercurrents}, we understand that the second term in the r.h.s. of Eq.~(\ref{Lambda_phi}) can be safely neglected unless we deliberately want to deal with an elongated system to probe the statistics of supercurrent states via the dependence of $\Lambda^{(\varphi)}_x$ on the aspect ratio $L_x^2/V$. This circumstance is quite important because for the loop models, $\Lambda^{(\varphi)}_x$ proves to be a very natural observable---the variance of the loop winding numbers, see Eq.~(\ref{Pollock_Ceperley}) of the next section. 

\begin{figure}[t]
    \includegraphics[width=1.0\columnwidth]{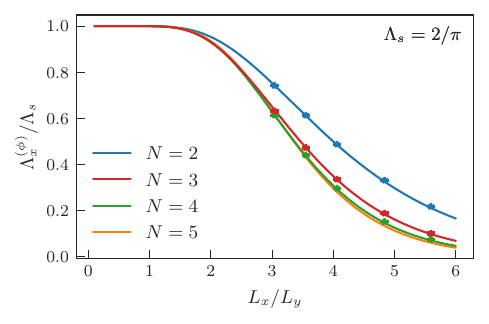}
    \caption{Phase twist response $\Lambda_x^{(\phi)}/\Lambda_s$ vs aspect ratio $L_x/L_y$ in two-dimensional $N$-component supercounterfluids at $\Lambda_s=2/\pi$.  Solid lines correspond to the effective theory, Eqs.~(\ref{Lambda_phi})--(\ref{av_def}). Dots represent numeric results obtained for minimal $N$-component loop models (see Fig.~\ref{fig:bond_conf} for $N=3$) based on relation (\ref{Pollock_Ceperley}) with $P$-values chosen so that $\Lambda_s = 2/\pi$ numerically for a square system. We set $L_x=49$, sufficient to render the finite-size effects negligibly small. 
    Note the absence of any fitting parameters. The condition $\Lambda_s=2/\pi$ is merely one of the reasonable choices of otherwise arbitrary values of models' parameter $P$.  Observe pronounced differences between the non-Borromean $N = 2$ case (equivalent to the single-component regime) and Borromean cases ($N \geq 3$).
    } \label{fig:C_err}
\end{figure}

Figure~\ref{fig:C_err} shows the response $\Lambda^{(\varphi)}_x$ in 2D supercounterfluids at $\Lambda_s=2/\pi$. The plot is instructive in two different ways. First, it demonstrates an impressive agreement---in the absence of any fitting parameters---between numeric data obtained for the minimal loop models, the $N$-component analogs of the model shown in Fig.~\ref{fig:bond_conf}, and the effective theory, Eqs.~(\ref{Lambda_phi})--(\ref{av_def}). Second, it reveals a pronounced difference between the non-Borromean case $N=2$ (equivalent to the single-component case) and Borromean $N\geq3$ cases. 

\subsection{Pollock-Ceperley relation} 
\label{subsec:Pollock_Ceperley}

The system's response to the phase twist applied to the $\alpha$-th component is readily measured by the worm algorithm.
In a lattice system, the twist has the effect of adding an extra term to the phase differences in Eq.~(\ref{lat_grad})
for the $\alpha$-th component between any two neighbouring sites in the $x$-direction:
\be
\nabla \theta_\alpha^{(b)} \, =\, \theta_{\alpha s_2} - \theta_{\alpha s_1} \, \to  \, \nabla \theta_\alpha^{(b)} + \varphi_0/L_x,
\label{lat_grad_twist}
\ee
where we assume that $x_{s_2} > x_{s_1}$; if the opposite is true, we flip the sign of $\varphi_0 /L_x$.
For each such bond the factor $Q$ gets modified to
\be
Q\left( \{ \nabla \theta\} \right) \, \to\,  \sum_{\bf n} F_{\bf n} \exp \left( i \sum_{\beta = 1}^N n_\beta \nabla \theta_\beta \right)\exp{\left(i\frac{n_\alpha\varphi_0}{L_x}\right)}  \, .
\label{Q_Fourier_twist}
\ee

For any given loop configuration $\{j\}$ the contribution to the partition function will thus be modified by the factor $e^{i\varphi_0 W_{\alpha,x}}$,
where $W_{\alpha,x} = L_x^{-1}\sum_{b_x}j_\alpha^{(b_x)}$ and $b_x$ denotes all bonds in the $x$-direction.
The quantity $W_{\alpha,x}$ is the total loop winding of the $\alpha$-th component in the $x$-direction
and is easily extractable during simulations with the worm algorithm.
By defining $Z_{W_{\alpha,x}}$ to be the total weight of all configurations with winding $W_{\alpha,x}$ at zero twist 
we see that the partition function can be written as
\begin{equation}
    Z = \sum_{W_{\alpha,x}}Z_{W_{\alpha,x}}e^{i\varphi_0 W_{\alpha,x}}.
\end{equation}

We see that
\begin{equation}
    -\frac{1}{Z} \left. \frac{\partial^2 Z}{\partial\varphi_0^2}\right|_{\varphi_0 = 0} = \frac{1}{Z}\sum_{W_{\alpha,x}}Z_{W_{\alpha,x}}W_{\alpha,x}^2 = \big\langle W_{\alpha,x}^2 \big\rangle \, , 
\end{equation}
where $W_\alpha$ is the loop winding of the $\alpha$-th component.
This leads to the Pollock-Ceperley relation 
\be 
\Lambda^{(\varphi)}_x \, =\,  {L_x^2 \over V } \, \langle W^2 \rangle \, ,
\qquad \langle W^2 \rangle \, =\, N^{-1} \sum_{\alpha=1}^{N} \, \langle W_\alpha^2 \rangle \, ,\quad
\label{Pollock_Ceperley}
\ee
where we also take into account that for better statistics one should average loop winding variances over all the components.

\section{Fluctuations of the Phase Fields in a Borromean Supercountrerfluid}
\label{sec:phase_field_fluct}

\subsection{Single-component case} 

We will see that the Borromean case can be reduced to a certain superposition of independent single-component cases. Therefore, we start with reviewing the single-component case (see, {\it e.g.}, \cite{sbpbook}). Here the fluctuations of the field of the phase, $\varphi(\bf r)$,  are described by the partition 
\begin{equation}
Z =\int   e^{-A[\varphi]} \, {\cal D} \varphi  \, , \qquad A[\varphi] = \int {\cal A}\, d^d r \, , \label{partit}
\end{equation}
\begin{equation}
 {\cal A} =  \frac{\Lambda_s}{2} \ (\nabla \varphi)^2 \, . \label{A}
\end{equation}
The field $\varphi$ varies significantly only over large enough
distances. Mathematically, this fact can be expressed by introducing a
UV cutoff, $k_*$, for the Fourier harmonics of $\varphi$ (system volume is set equal to unity):
\begin{equation}
\varphi ({\bf r}) \, =\,  \sum_{k < k_*} \varphi_{\bf k} \, e^{i{\bf k}
\cdot {\bf r}} .
\label{cutoff}
\end{equation}
Apart from the constraint
\begin{equation}
\varphi_{-{\bf k}} = \varphi_{\bf k}^*  ,  \label{constr}
\end{equation}
necessary for $\varphi$ to be real, all the Fourier harmonics fluctuate independently:
\begin{equation}
Z\, \propto\,  \tilde{\prod_{\bf k}}  \, Z_{\bf k}\,  ,  \label{indip}
\end{equation}
\begin{equation}
Z_{\bf k} = \int e^{-\Lambda_s k^2 
|\varphi_{\bf k}|^2} d \varphi_{\bf k}\,  . \label{Z_k}
\end{equation}
The tilde sign on the product means that, under the constraint  (\ref{constr}),
the subscript ${\bf k}$ simultaneously stands for a given wavevector and its counterpart of the opposite direction
so that each factor in the product is statistically independent. This is also the reason why there
is no factor of 1/2 in the exponent in (\ref{Z_k}).

Due to the bilinearity of (\ref{A}), the correlation functions of the field $\varphi$ are expressed, by Wick's theorem, in terms of sums of products of the pair
correlator (here, we also take into account the translation invariance of the problem)
\begin{equation}
G({\bf r})\,  = \, \langle \, \varphi({\bf r}') \, \varphi ({\bf r}'+{\bf r}) \, \rangle \, =
  \,  \langle \, \varphi(0) \, \varphi ({\bf r})\,  \rangle \,  .
\label{G_r}
\end{equation}
Taking Fourier transform and utilizing (\ref{Z_k}), one obtains
\begin{equation}
G({\bf r})\,  =\,  \sum_{k < k_*} G_{\bf k}  \, e^{i{\bf k} \cdot {\bf
r} } \, , \quad   G_{\bf k} = \langle \, |\varphi_{\bf k}|^2
\rangle = \frac{1}{\Lambda_sk^2}\, . \label{G_k}
\end{equation}

The simplest  off-diagonal correlator demonstration characteristic long-range behavior is the single-particle density matrix:
\begin{equation}
\rho ({\bf r}) \propto \langle e^{i \varphi ({\bf r}) - i\varphi
(0)} \rangle \,  , \qquad  \qquad r \to \infty\,   . \label{rho}
\end{equation}
The average (\ref{rho}) is readily evaluated by the formula
\begin{equation}
\langle e^{iQ} \rangle = e^{-\langle Q^2 \rangle /2} \, ,
\label{form}
\end{equation}
valid for any Gaussian quantity $Q$,  by Wick's
theorem. Substituting $[\varphi ({\bf r}) - \varphi (0) ]$ for $Q$ in Eq.~(\ref{form}), one finds (for $r \to \infty $)
\begin{equation}
\rho(r)\, \to \, n_{k_*} \exp \left[ - \int_{k<k_*} \!\!\!\!(1-\cos {\bf k} \cdot
{\bf r}  ) \, G_{\bf k}\,  {d^d k \over (2\pi )^d }
 \right]   .
\label{asymp2}
\end{equation}
Here, the summation over momenta is replaced with integration. The prefactor $n_{k_*}$ depends on the
choice of the cutoff momentum $k_*$ in such a way that the product of $n_{k_*}$ and the
exponential are $k_*$-independent.

In 3D,  $ n_{k_*} \to n_0$ at $k_* \to 0$, where $n_0$ is a finite constant---the condensate density. 
Correspondingly, the integral in the exponent
(\ref{asymp2}) becomes arbitrarily small as $k_*\to 0$,
so that one can expand the exponential and arrive at Bogoliubov's formula for the Fourier transform of $\rho(r)$:
\begin{equation}
\rho_{\bf k}\,  \to\,  (2\pi)^3 n_0\,  \delta({\bf k}) + \frac{n_0}{\Lambda_sk^2} \,
 , \quad  k \to 0\quad   (d=3)\,  . 
 \label{rho_k}
\end{equation}
In the real space, this implies
\begin{equation}
\rho(r)\,  \to\,   n_0\,\left( 1 + \frac{1}{4\pi \Lambda_s\,  r} \right) \,
 , \quad  r \to \infty \quad   (d=3)\,  . \label{rho_r_3D}
\end{equation}

In 2D, the integral in (\ref{asymp2}) behaves like  $(2\pi \Lambda_s)^{-1}
\ln (k_* r) $, at $r \gg 1/k_*$. As a result, the main
contribution to the integral comes from the interval between two cutoffs
$ 1/r < k <k_*$ where the function under the integral is $\propto 1/k^2$.
This leads to the logarithmic behavior.
Hence, in a 2D case, there is a power-law decay of the single-particle density matrix:
\begin{equation}
\rho(r) \propto 1/r^{\gamma}\,  ,\qquad   r \to \infty
\qquad  (d=2)\,  , 
\label{power_dec}
\end{equation}
with the exponent $\gamma$ given by
\begin{equation}
\gamma = {1\over 2\pi \Lambda_s }\,  . \label{expon}
\end{equation}
From this expression, we see that $n_{k_*}$ in Eq.~(\ref{asymp2}) should vanish as $n_{k_*} \propto k_*^{\gamma}$ when $k_*$ is decreased to ensure the physical answer remains cutoff independent.

\subsection{Borromean case} 

Our analysis will be based on the distribution (\ref{tilde_partit_B})--(\ref{tilde_A_B}).
The correlator of our interest is the composite density matrix
\begin{equation}
\rho ({\bf r}) \propto \langle e^{i \phi_\alpha ({\bf r}) - i\phi_\alpha
(0)} \rangle \,  = \,  e^{-{\cal K} ({\bf r})} \, , \label{rho_B}
\end{equation}
\begin{equation}
{\cal K} ({\bf r}) = \langle  \phi_\alpha(0) \phi_\alpha(0) -   \phi_\alpha({\bf r}) \phi_\alpha(0)\rangle \, . \label{K_B}
\end{equation}
By symmetry between different components,  correlator ${\cal K}$ does not depend on $\alpha$. This allows us to write it in the following convenient form
\begin{equation}
{\cal K} ({\bf r}) = {1\over N-1 }\left \langle  \sum_{\alpha=1}^{N-1 }\,  [\phi_\alpha(0) \phi_\alpha(0) -   \phi_\alpha({\bf r}) \phi_\alpha(0) ] \right\rangle \, , \label{K_B_sym}
\end{equation}
which is invariant with respect to the unitary transformation of the fields,
\be
\phi_\alpha = \sum_{\nu =1}^{N-1} U_{\alpha \nu} \varphi_\nu.
\label{unitary}
\ee
Therefore, the problem reduces to diagonalizing the free-energy density (\ref{tilde_A_B}) by certain transformation (\ref{unitary}), calculating independent single-component correlators
\be
{\cal K}_\nu ({\bf r}) = \langle  \varphi_\nu(0) \varphi_\nu(0) -   \varphi_\nu({\bf r}) \varphi_\nu (0)\rangle \, ,
\label{K_nu}
\ee
and summing them up:
\be
{\cal K} ({\bf r}) = {1\over N-1 } \sum_{\nu=1}^{N-1} {\cal K}_\nu ({\bf r}) \, .
\label{K_B_result}
\ee
Furthermore, we do not need to know the explicit form of the unitary matrix $U_{\alpha \nu} $ because the value of the correlator (\ref{K_nu}) is controlled only by the corresponding eigenvalue of the diagonalized free-energy density (\ref{tilde_A_B}). Finally, finding these eigenvalues is straightforward because the first term in (\ref{tilde_A_B}) is invariant with respect to the unitary transformations, and the second term is given by the standard all-ones matrix. The all-ones matrix has only one nonzero (non-degenerate) eigenvalue that is equal to the rank of the matrix; all other eigenvalues are identically equal to zero. This brings us to the following single-component free-energy densities
\be
{\tilde A}_\nu = {N\Lambda_s\over 2(N-1)} (\nabla \varphi_\nu)^2 \, , \quad  \nu \neq  \nu_* \, , \quad  \nu_*= N-1\, , 
\label{general}
\ee
\be
{\tilde A}_{\nu_*} = {\Lambda_s \over 2(N-1)} (\nabla \varphi_{\nu_*})^2 \, , \qquad   \nu_*=N-1\, . 
\label{special}
\ee
Using the single-component results, we get 
\begin{equation}
\rho(r)\, \to \, n_{k_*} \exp \left[ - \int_{k<k_*} \!\!\!\!(1-\cos {\bf k} \cdot
{\bf r}  ) \, G_{\bf k}\,  {d^d k \over (2\pi )^d }
 \right]   ,
\label{asymp_B}
\end{equation}
where
\[
G_{\bf k} = {1\over N-1} \sum_{\nu = 1}^{N-1} G_{\bf k}^{(\nu)} \, ,
\]
\be
G_{\bf k}^{(\nu)} = {(N-1)\over N\Lambda_s k^2} \, , \quad  \nu \neq  \nu_* \, , \quad  \nu_*= N-1\, , 
\label{G_general}
\ee
\be
G_{\bf k}^{(\nu_*)} = {(N-1)\over \Lambda_s k^2} \, , \qquad   \nu_*=N-1\, . 
\label{G_special}
\ee
With Eqs.~(\ref{G_general}) and (\ref{G_special}) we find

\be
G_{\bf k} = {2(N-1)\over N \Lambda_s  k^2} \, ,
\label{G_Brm_Lambda_a}
\ee
and comparing it to $G_{\bf k}$ of Eq.~(\ref{G_k}), we conclude that counterflow analogs of Eqs.~(\ref{rho_k})--(\ref{expon}) are obtained by the
replacements
\be
\Lambda_s \, \to \, {N\over 2(N-1) }  \, \Lambda_s\, ,
\label{replacement}
\ee
meaning that
\be
\gamma \, =\ {N-1\over \pi N \Lambda_s} \, .
\label{gamma}
\ee
The largest possible value of $\gamma$ takes place at the BKT point. According to Eq.~(\ref{NK_relation}), this value is given by
\be
\gamma^{\rm (BKT)}(N) \, =\, {N-1\over 2 N } \, .
\label{gamma_BKT}
\ee
Consistent with the isomorphism between the two-component supercounterfluid and single-component superfluid,  $\gamma^{\rm (BKT)}(2)=1/4$. The inequality $\gamma^{\rm (BKT)} (N) > 1/4$  thus can be viewed as a Borromean feature.

As we explained in Sec.~\ref{sunsec:Numeric_protocol}, we simulate the composite density matrix of the minimal loop model by the worm algorithm. An optimal way of clearly revealing the power-law behavior of $\rho_{s_1 s_2}$ (\ref{rho_Bor_latt})---allowing us to practically eliminate the finite-size systematic errors while substantially reducing statistical noise---is to work with the following averaged quantity:
\be
\rho_{\rm avg}(L/2) \, =\, C_{\rm norm}(L/2) \sum_{s_1 s_2} \, \rho_{s_1 s_2} \theta_{s_1 s_2}^{(L/2)}  \, ,
\label{rho_av}
\ee
\be
\theta_{s_1 s_2}^{(L/2)}\, =\, \left\{\begin{array}{c} 1\, , ~~ \mbox{if} ~~ r_{s_1 s_2} \in [0.4L, 0.5L] \, ,\\~ 0\, ,\,  ~\mbox{otherwise}\, ,\qquad \qquad  \qquad \qquad  \end{array}\right.
\label{theta_s1_s2}
\ee
where $r_{s_1 s_2}$ is the distance between the sites $s_1$ and $s_2$ and $C_{\rm norm}$ is the normalization constant:
\be
C_{\rm norm}^{-1}(L/2) \, =\,  \sum_{s_1 s_2} \,  \theta_{s_1 s_2}^{(L/2)} \, .
\label{C_norm}
\ee
As a function of system size $L$, the averaged composite density matrix scales as 
\be
\rho_{\rm avg}(L/2) \, \propto\, 1/L^{\gamma} \, ,
\label{rho_avg_scaling}
\ee
precisely in the same way as the off-diagonal correlator $\rho(L/2)$.
Using $\rho(L/2)$ directly however is not optimal since we would then only use the simulated data for a thin line of separations
exactly at the distance $L/2$.
By contrast many more separations (which we anyways sample in our simulations) enter into the calculation for $\rho_{\text{avg}}(L/2)$, making it a much better choice of observable. 

\begin{figure}[t]
    \includegraphics[width=1.0\columnwidth]{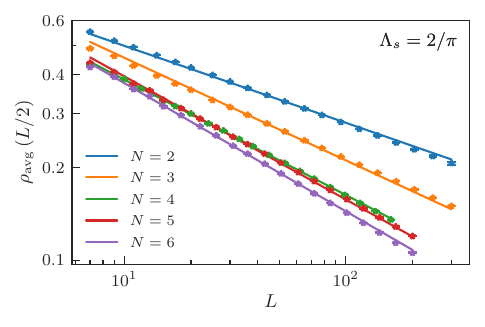}
    \caption{Averaged composite density matrix $\rho_{\text{avg}}\left(L/2\right)$ vs the system size $L$ for different numbers of components in ordered phase close to the transition point. 
    Numeric data are fitted as $\rho_{\text{avg}}(L/2)=A r^{-\gamma}$, with $\gamma = \gamma^{\rm (BKT)}(N)$; see Eq.~(\ref{gamma_BKT}). Note excellent agreement between the numeric data and basic theory that does not take into account the renormalization flow of superfluid stiffness $\Lambda_s \equiv \Lambda_s(L)$ at the BKT point.}
    \label{fig:cdm_poly}
\end{figure}

The scaling of $\rho_{\text{avg}}(L/2)$, also demonstrating the validity of relation (\ref{gamma_BKT}), is seen in Fig.~\ref{fig:cdm_poly}.

\section{Borromean BKT-type transition}
\label{sec:BKT}

\subsection{Basic relations}

Our goal here is to generalize the Kosterlitz-Thouless renormalization-group (RG) theory. In doing so, we follow the top-down approach described, {\it e.g.}, in Ref.~\cite{sbpbook}. The central role will be played by the relation between the parameter $\Lambda_s$ and the response to the phase twist discussed in Sec.~\ref{subsec:resp_phase_twist}. As opposed to the main goal of that section---tracing the statistical effect of supercurrent states---here we are not interested in this effect and deal exclusively with the topological sector $\Sigma({\bf M})=0$. What we want to trace here is how $\Lambda_s$ ``flows"  with the system size, the effect being due to the contributions from the vortex-antivortex pairs of the size $\sim L$.  In the absence of contribution from the supercurrent states, Eqs.~(\ref{Lambda_phi_def})--(\ref{Lambda_phi}) yield
\be
\Lambda_s \, =\, {L_x^2 \over V } \left. {\partial^2  F \over \partial \varphi_0^2} \right\vert_{\varphi_0=0} \, , \qquad F \, =\, - \ln Z \, .
\label{Lambda_by_phase_twist}
\ee
We remind that here $\varphi_0$ is the phase twist applied to a certain component $\alpha$. To trace the dependence of $\Lambda_s$ on the system size, we need to explore the $\varphi_0$-dependent  contributions of the vortex pairs of the size $\sim L$ to the free energy $F$. 

In the Borromean system, vortex-antivortex pairs in any component couple to the persistent current created by the phase twist. Therefore, the renormalization of $\Lambda_s$ by vortex-antivortex pairs has a pronounced ``Borromean" nature. In this sense, the fact that the flow equations for $\Lambda_s(L)$, which we derive below, have precisely the same mathematical form as in a single-component case should not be considered trivial. Furthermore, direct access to the concentration of vortex--antivortex pairs would allow one to reveal the Borromean aspect of the RG flow explicitly.

From Eq.~(\ref{eq:pair_energy}) we see that a vortex-antivortex pair of the same component in the absence of phase twists has the energy (up to irrelevant constant term)
\be
E_{\rm pair}(\Rv) \, =\,  2\pi \Lambda_s\ln{R} \, , 
\ee
where $\Rv$ is the separation vector between the vortices.
In the presence of a phase twist $\varphi_0$ in component $\alpha$ in the $x$-direction, this energy gets modified to 
\be
E_{\rm pair}^{(\alpha)}(\Rv, \varphi_0) \, =\, E_{\rm pair}(\Rv)  - \frac{2\pi \Lambda_s \varphi_0}{L_x} \, \hat{x}\cdot(\zh\times\Rv) \, ,
\label{coupling_same}
\ee
\be
E_{\rm pair}^{(\beta\neq \alpha)}(\Rv, \varphi_0) \, =\,  E_{\rm pair}(\Rv)\,  +\, \frac{2\pi \Lambda_s \varphi_0}{(N\! - \! 1)L_x} \, \hat{x}\cdot(\zh\times\Rv) \, .
\label{coupling_different}
\ee
Note that vortices in other components $\beta\neq\alpha$ do couple to the phase twist in component $\alpha$,
with an important reminder that such vortices exist only at $N\geq 3$ so that we are dealing with a Borromean effect. In terms of the sign and relative strength of these couplings, there is a close analogy with Eqs.~(\ref{eq:pair_energy}) and (\ref{eq:diff_pair_energy}). In the context of the $N=2$ case, it is instructive to observe that here both (\ref{coupling_same}) and (\ref{coupling_different}) still apply, but describe the same setup---in two different gauges.

\subsection{Renormalization-group flow}

Within the (asymptotically exact) top-down approach, we are interested only in the large scales of distances, $\sim L$,  and correspondingly, large pairs, $R \sim L$. The gas of such pairs is asymptotically dilute, and we can express the free energy using only the partition function for isolated vortex-antivortex pairs.
The free energy of a system of size $L$ can therefore be written as
\bea
    F(L) = \left(\frac{L}{l}\right)^2F(l) - V \int_{l\leq|\Rv|\leq  L}\kern-0.7cm\upd^2R \, e^{-E_{\rm pair}^{(\alpha)}(\Rv, \varphi_0)} \nonumber \\
    - V\sum_{\beta \neq \alpha }\int_{l\leq|\Rv|\leq  L}\kern-0.7cm\upd^2R \,  e^{-E_{\rm pair}^{(\beta\neq \alpha)}(\Rv, \varphi_0) } \, , ~~~
    \label{eq:feL}
\eea
where $F(l)$ is the free energy of a system of the size $l < L$ and we assume that the value of $\Lambda_s$ in the pair energies (\ref{coupling_same})--(\ref{coupling_different}) equals $\Lambda_s(l)$.

Applying relation (\ref{Lambda_by_phase_twist}) to Eq.~(\ref{eq:feL}) we get
\be
\Lambda_s(L) = \Lambda_s(l) - 4\pi^3\Lambda_s^2(l)\frac{N}{N-1}\int_{l}^L\upd R\hspace{5pt}R^3e^{-E_{\rm pair}(R)}\, . \label{eq:lambda_scale}
\ee
The origin of the factor $N/(N-1)$ is quite instructive:
\be
{N\over N-1} \, =\, 1 \, +\,  {1\over N-1} \, .
\label{N_over_N_minus_1}
\ee
The first term in the r.h.s. comes from the first integral in (\ref{eq:feL}). It is responsible for the renormalization of a superflow created by the phase twist in component $\alpha$ by the vortex-antivortex pairs of the same component. Therefore, this term corresponds to the single-component physics. The second term originates from the second integral in (\ref{eq:feL}). As we already mentioned, the $N=2$ case is special. Here, the second integral simply double-counts the vortex pairs in the same way as the first one. Therefore, we should exclude the second term (or, equivalently,  divide the expression by the factor of 2), thereby getting the result identical to that for the single-component superfluid. Hence, the second term is a signature of Borromean physics. It has the largest value at $N=3$ and vanishes as $N\to \infty$, in which limit the renormalization of the superfluid stiffness for each component $\alpha$ takes place independently of the other components, effectively reproducing the single-component regime.

Apart from the Borromean factor $N/(N-1)$, relation (\ref{eq:lambda_scale}) proves to be the same as in a single-component superfluid, cf.~\cite{sbpbook}. This immediately allows us to re-use the single-component results. Indeed, recalling that the constant factor in front of the integral in the r.h.s. of (\ref{eq:lambda_scale}) plays only a minor role being absorbed into the definition of non-dimensionalized density of vortex pairs \cite{sbpbook}, we conclude that the resulting RG equations can be cast into the same form:
\be
{d w \over d \ln (l/l_0) } = -2 g \, ,
\label{KT1}
\ee
\be
{d g \over d \ln (l/l_0)} = -2 w g \, ,
\label{KT2} 
\ee
where $l_0$ is a certain ultraviolet length cutoff,
\be
w(l) = \pi\Lambda_s(l) - 2 \, .
\label{w}
\ee
Finally,  $g(l)$ is the rescaled (nondimentionalized) density of the vortex pairs of the size $\sim l$, which differs from the standard single-component definition (see \cite{sbpbook}) by the Borromean factor:
\be
g(l)\, \to \, {N\over N-1} \, g(l) \, .
\label{g_Borromean}
\ee
Since RG equations (\ref{KT1})--(\ref{KT2}) have the same mathematical form as in the single-component case, we can simply quote the single-component solutions (see, {\it e.g.} \cite{sbpbook}). For the flow of superfluid stiffness in the vicinity of the BKT point, we have
\bea
w(l) \, &=&\, {\sqrt{C}\over \tanh \left[\sqrt{C} \, \ln(l/l_0)\right]} \quad (C > 0)\, , \label{positive_C}\\
w(l) \, &=& \, {\sqrt{|C|}\over \tan \left[\sqrt{|C|} \, \ln(l/l_0)\right]}  \quad (C < 0) \, ,\label{negative_C} \\
w(l) \, &=&\, {1\over \ln(l/l_0)} \qquad \qquad \qquad (C = 0)\, ,\label{zero_C}
\eea
where $C$ is an analytic function of control parameter(s) of the theory, with $C=0$ corresponding to the BKT point. The function $g(l)$ is then given by
\be
g(l) \, =\, w^2(l) - C \, .
\label{g_L}
\ee

Along similar lines (c.f. \cite{sbpbook}), we produce the expression for the reduced density matrix $\rho(r)$ starting with the differential version of Eqs.~(\ref{power_dec})--(\ref{expon}),
\be
{d \ln \rho (r) \over d\ln (r/l_0)} \, = \, - {N-1\over \pi N \Lambda_s (r)} \, ,
\label{rho_flow_1}
\ee
which takes into account the slow flow of $\Lambda_s$ with the scale of distance. Expressing $\Lambda_s(r)$ in terms of $w(r)$ in accordance with (\ref{w}) and expanding up to the leading correction in terms of the small parameter $w\ll 1$, we get
\be
{d \ln \rho (r) \over d\ln (r/l_0)} \, = \, - \gamma_0 + {\gamma_0 \over 2}w(r) \, ,
\label{rho_flow_2}
\ee
\be
\gamma_0\, \equiv \, \gamma_{\rm BKT} \, =\, {N-1\over 2N} \, .
\label{gamma_0}
\ee
Finally, using explicit expressions (\ref{positive_C})--(\ref{zero_C}) for $w$, we perform integration and get
\begin{align}
    &\rho(r) = \frac{B}{r^{\gamma_0}}\left[\frac{1}{\sqrt{C}}\sinh\left(\sqrt{C}\ln{\frac{r}{l_0}}\right)\right]^{\gamma_0/2} ~ &(C > 0) , \label{rho_positive_C}\\
    &\rho(r) = \frac{B}{r^{\gamma_0}}\left[\frac{1}{\sqrt{|C|}}\sin\left(\sqrt{|C|}\ln{\frac{r}{l_0}}\right)\right]^{\gamma_0/2} ~ &(C < 0) , \\
    &\rho(r) = \frac{B}{r^{\gamma_0}}\left[\ln{\frac{r}{l_0}}\right]^{\gamma_0/2} ~ &(C = 0) , \label{rho_zero_C}
\end{align}
where $B$ is a certain $P$-indepedent constant.  

In conclusion of this section, it is worth observing that solution (\ref{g_L}) does not allow us to see the Borromean nature of the RG flow unless we have direct access to the concentration of vortex pairs and can separate $g(l)$ into two factors: the pure concentration of vortex pairs and the Borromean factor $N/(N-1)$, see Eq.~(\ref{g_Borromean}).

\subsection{Numeric data vs effective long-wave description}

It is very instructive to compare the results of numeric simulations of the 2D minimal Borromean ($N=3$) loop model with the predictions of the effective long-wave theory. 

\begin{figure}[t]
    \includegraphics[width=1.0\columnwidth]{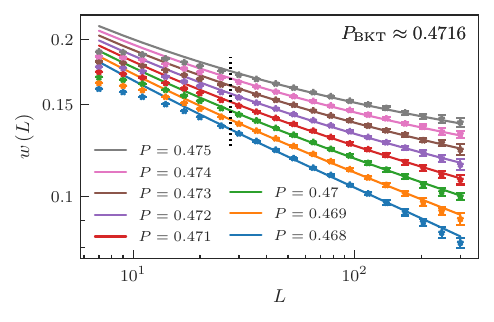}
    \caption{Reduced winding number $w\left(L\right)$ vs the system size $L$ in a 2D three-component system in the vicinity of the BKT transition.  
    The lines correspond to Eq.~(\ref{w_expanded}) with free parameters obtained by the joint fitting procedure as explained in the text. The vertical dashed line cuts off the low-$L$ data points excluded from the fitting protocol's objective function. All the data for $P<0.47$ are also excluded from the objective function.} \label{fig:rgf}
\end{figure}
\begin{figure}[t]
    \includegraphics[width=1.0\columnwidth]{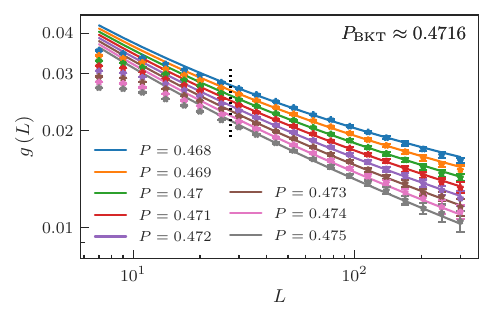}
    \caption{Rescaled density of the vortex pairs $g\left(L\right)$ vs the system size $L$ in a 2D three-component system in the vicinity of the BKT transition. Numeric data and analytical curves correspond to those shown in Fig.~\ref{fig:rgf} under transformation $g\left(L\right) = w^2(L)-C$; see Eq.~\eqref{g_L}.} \label{fig:g_dep}
\end{figure}
\begin{figure}[t]
    \includegraphics[width=1.0\columnwidth]{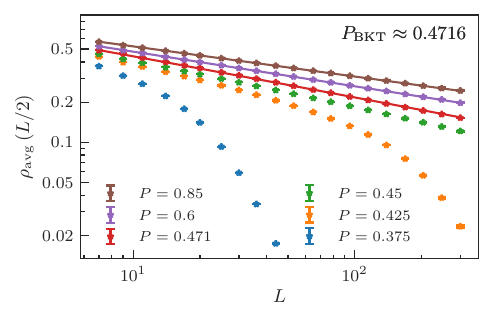}
    \caption{Composite density matrix $\rho_{\rm avg}\left(L/2\right)$ vs the system size $L$ in a 2D three-component system. Symbols with error bars show the numeric data. Results for $P=0.471$ are approximated using (\ref{rho_avg_flow}), with free parameters obtained from the joint fitting procedure described in the text. Results for $P=0.6$ and $P=0.85$ are fitted as $\rho_{\text{avg}}(L/2)=A r^{-\gamma}$, with the exponent $\gamma$ defined in \eqref{gamma}.  
 In the disordered phase, correlations decay exponentially at large enough distances, as demonstrated by the data for $P=0.375$ and $P=0.425$. At $P=0.45$, the exponential decay takes place at distances much larger than the size of our simulation box. }
    
    \label{fig:cdm}
\end{figure}

Given that RG theory is controlled only in the close vicinity of the BKT point, it is sufficient to treat $C$ as a linear function of the parameter $P$:
\be
C \, =\,C_0 \, (P - P_{\rm BKT}) \, ,
\label{C_of_P}
\ee
where $C_0$ is a certain positive constant and $P_{\rm BKT}$ is the critical value of $P$. For the same reason, we can treat $l_0$ as a $P$-independent constant. Also, taking into account that the $L$ dependence is logarithmically slow while the controlled long-wave description requires $|C|\ll 1$, for our practical purposes, it would be safe to Taylor-expand Eqs.~(\ref{positive_C})--(\ref{zero_C}) up to the first subleading term:
\be
w(L) \, \approx \, {1\over \ln(L/l_0)}\left[ 1 + {C\over 3}\ln^2 \left( {L\over l_0} \right) \right] . 
\label{w_expanded}
\ee

From the RG theory we can also find a similar expression for $\rho_{\rm avg}(L/2)$ by Taylor expanding Eqs.~(\ref{rho_positive_C})--(\ref{rho_zero_C}). As opposed to $w(L)$ the result depends on $N$; for $N=3$ we have
\be
\!\! \rho_{\rm avg}(L/2) \approx \frac{B}{L^{1/3}}\left[\ln\left(L/l_{0}\right)\right]^{1/6}\left[1\! +\! \frac{C}{36}\ln^{2}\left(\frac{L}{l_{0}}\right)\right]\!  .
\label{rho_avg_flow}
\ee
Note that the relative value of the subleading term in (\ref{rho_avg_flow}) is an order of magnitude smaller than in (\ref{w_expanded}).

The four parameters, $P_{\rm BKT}$, $C_0$, $l_0$, and $B$, were obtained by jointly fitting numeric data for $w(L)$ and $\rho_{\rm avg}(L/2)$ with the functions (\ref{w_expanded}) and (\ref{rho_avg_flow}), respectively, in close proximity to the phase transition: $P\in [0.470, \, 0.475]$.
We found that $P_{\rm BKT}=0.4716\pm0.0003$, $C_0=2.61\pm0.15$, $\ln l_0=-3.11\pm 0.11$, $B=0.717\pm0.003$. Additional fitting, performed for $P$-dependent functions $B$ and $l_0$, showed that even in this case best estimations for these parameters are $P$-independent constants. 

Figure~\ref{fig:rgf} demonstrated perfect agreement between numeric data and asymptotic analytic predictions (\ref{positive_C})--(\ref{w_expanded}). Also instructive, the agreement starts at rather large system sizes $L\sim 30$. This is not particularly surprising given that the Kosterlitz-Thouless renormalization flow requires the system to accommodate a pair of well-separated vortices. 

The conditions of applicability of RG equations (\ref{KT1})--(\ref{KT2}) are $w\ll 1$ and $g \ll 1$ \cite{sbpbook}. The data of Fig.~\ref{fig:rgf} demonstrate consistency with the former requirement; from Fig.~\ref{fig:g_dep} we see that the latter inequality is also satisfied. Still, in Fig.~\ref{fig:rgf}, one can notice that the $L > 200$ data for $P<0.47$ starts to develop tiny deviations from Eq.~(\ref{w_expanded}). This is likely to manifest the subleading correction to the RG flow, which is of substantial fundamental interest, as discussed in Sec.~\ref{sec:discussion}. The quantitative theory of this correction goes beyond the scope of the present paper.

In contrast to $w(L)$, the critical flow for $\rho_{\rm avg}(L/2)$ is perfectly fitted by Eq.~(\ref{rho_avg_flow}) starting with significantly smaller distances; see the $P=0.471$ line in Fig.~\ref{fig:cdm}. This is because the effect of renormalization of $\Lambda_s(L)$ on $\rho_{\rm avg}(L/2)$ is relatively weak, as is evident from  Eq.~(\ref{rho_avg_flow}) and as could also be expected based on the data shown in Fig.~\ref{fig:cdm_poly}, where the fitting curves simply neglect the flow of $\Lambda_s$ with $L$.

\section{Discussion}
\label{sec:discussion}

Borromean supercounterfluids feature correlation and topological properties distinguishing them from their single- and two-component counterparts. Especially interesting is the component-symmetric case characterized by a distinctively different universality class of the supercounterfluid-to-normal phase transition. A natural way of describing universal long-wave properties of Borromean supercounterfluids is in terms of compact-gauge-invariant effective action democratically treating all the $N$ components of the system. 

Compact-gauge-invariant theory substantially differs from its usual local U(1) counterpart. In the latter case, slight variations in phase fields can be ``undone" by gauge field at no energy cost, while singular configurations---vortices---are robust with respect to gauge transformation. The compact-gauge invariance---an intrinsic property of Borromean systems that does not involve gauge fields---allows to gauge in and out special vortex configurations in which all the $N$ components have the same phase winding. 

Using compact-gauge invariance as the guiding principle, we formulated long-wave effective action.
Also, based on the principle of compact-gauge invariance, we formulated two classes of models of Borromean counterfluids and established an explicit relationship (duality) between them. The first class is represented by the Borromean generalization of the XY model, Eq.~\eqref{H_Bor}. The second class consists of the loop models, of which particularly simple and elegant is the minimal model illustrated in Fig.~\ref{fig:bond_conf}.
Loop representation is free of gauge redundancy while being explicitly component-symmetric.
Even more importantly, loop models are perfectly suited for the simulations with worm algorithm;
all our numeric data were obtained this way.

We investigated the system's off-diagonal correlations and response to the phase twist finding, which both reflect the Borromean nature of the system. Our analytic predictions are in perfect agreement with the numeric data.

In the 2D case, we also studied Berezinski-Kosterlitz-Thoules(BKT)-type critical behavior. Here, the renormalization of superfluid stiffness $\Lambda_s$ by vortex-antivortex pairs has a pronounced ``Borromean" character due to the interaction between the vortices of different components.
Remarkably, the resulting flow equations (\ref{KT1})--(\ref{KT2}) have the same mathematical form as in a single-component case, provided the Borromean factor $N/(N-1)$ originating from the intercomponent vortex interaction is absorbed into the renormalized density of vortex-antivortex pairs, see Eq.~(\ref{g_Borromean}).

 There are still exciting and fundamental questions to address in future studies.
 One intriguing way of observing the effect of modular arithmetic of topological charges---a signature Borromean feature---is by resolving subleading correction to the renormalization of the superfluid stiffness of the 2D three-component system in the vicinity of the BKT transition;
 desirably on the normal side, where the subleading effects get stronger while allowing for a controllably accurate description within a broad range of system sizes. The subleading correction comes from the topologically neutral minimal non-binary vortex clusters. The unique property of the three-component supercounterfluid is that the minimal non-binary topologically neutral vortex cluster consists of only three vortices. In all other cases of BKT-type transitions, the minimal non-binary topologically neutral vortex cluster consists of four vortices, thus implying a different type of subleading correction to the flow.

The efficiency of the loop models comes at a price. They do not provide direct access to the statistics of vortices and thus do not allow us to see the factor $N/(N-1)$ enhancing the strength of the renormalization flow at the BKT point. To observe the $N/(N-1)$-enhancement, one has to simulate Borromean XY model (\ref{H_Bor}).

A big open question is how the Borromean nature manifests itself in the 3D criticality. In a superfluid system with global symmetries, including paired superfluid, the general expectation is that the transition to a symmetric state is continuous \cite{Kuklov2006decoinfined}. However, indications of first-order transitions from counterflow superfluid to a normal state were reported \cite{weston2019phase}, which suggests that compact-gauge invariance is significant for critical behavior in three dimensions and requires further investigation.

Finally, a separate direction of further study can be on Borromean systems that break only discrete symmetries. A case of particular interest---because of its relevance to the experimental system \cite{Grinenko2021state}---is the explicit breaking of symmetry down to $Z_2$ time-reversal symmetry, which confines counterflow currents down to finite length scales.

\begin{acknowledgments}
We thank Albert Samoilenka for useful discussions.
This work was supported by 
the Swedish Research Council Grants  2018-03659, 2022-04763, by Olle Engkvists Stiftelse and the Wallenberg Initiative Materials Science for Sustainability (WISE) funded by the Knut and Alice Wallenberg Foundation. BS acknowledges support from the National Science Foundation under Grant DMR-2335904.
The computations were enabled by resources provided by the National Academic Infrastructure for Supercomputing in Sweden (NAISS), partially funded by the Swedish Research Council through grant agreement no. 2022-06725.
\end{acknowledgments}

\bibliography{borcrit}

\end{document}